\documentclass[preprint2]{aastex}
\begin{document}

\title{The differential energy distribution of 
the universal density profile of dark halo}
\author{C. Hanyu and A. Habe}
\affil{Division of Physics, Graduate School of Science,
Hokkaido University, Sapporo 060-0810, Japan}
\email{chiaki@astro1.sci.hokudai.ac.jp}
\email{habe@astro1.sci.hokudai.ac.jp}

\begin{abstract}
We study the differential energy distribution of dark matter halos, carrying out cosmological $N$-body simulation.
We give an analytical formula of the differential energy distribution of dark matter in the halos obtained by the numerical simulation.
From the analytical formula we reconstruct the density profile described by the Navarro,
Frenk, \& White (NFW) profile.
The NFW profile is consistent with the analytical formula of 
our fractional mass distribution. 
We find that a parameter in our analytical formula of 
differential energy distribution which is related with 
the slope of inner cusp of dark halo.
We obtain the distribution function for the NFW profile which has sharp
cut off at the high binding energy.
We discuss physical reason of form of the analytical formula.
\end{abstract}


\section{Introduction}
It is interesting to understand how the density profiles of 
galaxies and clusters of galaxies have formed.
Navarro, Frenk and White (1995, 1996, 1997; NFW) have shown 
in their N-body simulations of Cold Dark Matter (CDM) in 
the standard biased CDM and four power law spectra with indices 
$n = 0,-0.5,$ and $-1$, open CDM ($\Omega_0 =0.1$) with 
power-law spectra ($n=0$ and $-1$), 
and $\Lambda$CDM cosmology that 
density profiles of dark halos have an universal profile described as
\begin{eqnarray}
	\rho(r) \propto \frac{1}
		{\left(\frac{r}{r_s}\right) \left( 1 + \frac{r}{r_s} \right)^2}.
\end{eqnarray}
Several investigators have shown that 
the formula provides a good fit to their numerical results 
(Cole and Lacey 1996, Tormen, Bouchet \& White 1997,
Huss, Jain \& Steinmmetz 1999, Thomas {\it et al.} 1998).
However, some other simulations (Fukushige \& Makino 1997;
Moore et al. 1998; Okamoto \& Habe 1999, 2000) indicate 
that their density profiles have 
steeper inner cusp than the NFW profile.
Jing (2000) gives his numerical results that
steeper cusps of density profile are found in recent mergers and 
in dark halos with substructures, by 
a large set of high resolution cosmological simulations.

Subramanian, Cen and Ostriker (2000) discussed 
the general theoretical grounds of the density profiles.
They discussed that there is a possible connection between  
slope of inner region of dark halo and 
the formation epoch and proposed that
there is possible relation to  cosmological parameters.

However, physical reason of the NFW profile have not been explored, 
although it is very important to understand 
formation of galaxies and clusters of galaxies.

We study the differential energy distribution of the NFW profile, 
$d M/ d \varepsilon$, where $\varepsilon$ is the binding energy. 
The differential energy distribution was studied to understand virialization process of $N$- body system (van Albada 1982). 
Binney (1982) shows that $dM/d \varepsilon$ of 
the elliptical galaxies is approximated by the Boltzmann factor, if 
surface brightness of elliptical galaxies obeys the 
de Vaucoureurs' $r^{1/4}$ law and their mass-to-light ratio is constant.
It is interesting to study phase space structure of virialized objects 
with the NFW profile, 
since the differential energy distribution may give insight of the relaxation process (Binney \& Tremain 1987).

In this paper we calculate formation of clusters of galaxies 
by $N$-body simulation and we study 
the differential energy distribution of the clusters in our numerical results.
We find an analytical formula of the fractional mass distribution 
that is defined by the differential distribution divided by mass of the object,
 is fitted by an analytical formula.
We show that 
the NFW profile is consistent with our analytical formula of 
the fractional mass distribution. 
Using the iteration method, we construct the density profile 
from the analytical formula to show how 
the slope of the cusp of the density changes with
 parameters in our analytical formula.

In the next section, 
we illustrate our numerical method.
In the section 3, we present our numerical results.
We show the iteration method and give results by this method 
in the section 4, and we summarize and 
discuss our results in the section 5.

\section{Numerical simulation}
We use the Hofman and Ribak's (Hofman and Ribak 1991)
 procedure to set initial conditions 
 in order to have a massive dark halo 
near the centre of a simulation box. 
The cosmology is SCDM model (e.g. Davis {\it et al}. 1985)
($\Omega = 1$, $\sigma_8 = 0.67$ , 
$H_0 = 100h$ km s${}^{-1}$ Mpc${}^{-1}$,and $h = 0.5$).

Numerical simulations are carried out using GRAPESPH code.
GRAPE is a special purpose hardware to calculate gravitation 
between N-body particles (Sugimoto {\it et al}. 1990).
We combined Smoothed Particle Hydrodynamics 
(SPH) (Monaghan, 1992) with GRAPE.
We select massive halos of which mass is as large as 
that of cluster of galaxies 
and calculate their density profile and the fractional mass distribution.

Mass of a CDM particle and a SPH particle are 
$5.89 \times 10^{11} M_{\odot}$ and 
$3.10 \times 10^{10} M_{\odot}$, respectively.
Gravitational softening length is 100 kpc.
Both number of CDM and SPH particles are 29855, respectively.
Size of the simulation box is $80$ Mpc.

\section{Numerical results}

\subsection{Density distribution}
Before we show our numerical results, we summarize the NFW profile.
NFW proposed that the profile of the dark halo of cosmological object in their numerical results as
\begin{eqnarray}
	\frac{\rho(r)}{\rho_{cr}} = \frac{\delta_c}{\left(\frac{r}{r_s}\right)
	\left( 1 + \left( \frac{r}{r_s} \right)\right)^2}, \label{NFW}
\end{eqnarray}
where $r_s = r_{200}/c$, $c$ is a dimensionless parameter and $\rho_{cr} = (3 H_0^2)/(8 \pi G)$ is the critical density of the universe .
Since the mass of the halo is 
$M(r_{200}) = 200 \times 4 \pi /3 \rho_{cr} r_{200}^3$, 
there is a relation between $\delta_c$ and $c$ as 
\begin{eqnarray}
	\delta_c = \frac{200}{3} \frac{c^3}{[ \ln{(1+c) - c/(1+c)} ]},
\end{eqnarray}
where $M(r_{200})$ is the mass of which averaged density 
inside $r_{200}$ is $200$ times $\rho_{cr}$. 

This density profile has a cusp in the inner region, 
$\rho(r) \propto r^{-1}$.
$\rho(r) \propto r^{-3}$ 
in the outer region.
NFW97 showed that the parameter $c$ decreases with halo mass. 

\placetable{table1}

Table 1 is physical values of our simulated clusters.
The units of mass, and 
X-ray temperature are $10^{15} M_{\odot}$ and $10^8 $ K,
respectively.

\begin{figure}
\plotone{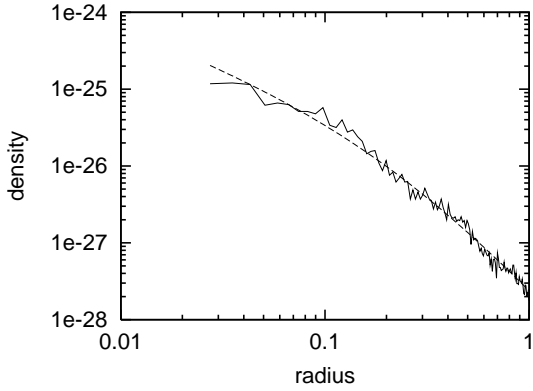}
\caption{Density profile of one of simulated clusters, CLc, and 
the NFW profile with $c = 4.4$. 
Solid line is our numerical result and dashed line is 
the NFW profile.
Radial distance is normalized by $r_{200}$. 
Our simulated clusters agree well with 
the NFW formula between the gravitational softening length to $r_{200}$.}
\label{typical_density_profile_with_NFW}
\end{figure}

\placefigure{typical_density_profile_with_NFW}

Figure \ref{typical_density_profile_with_NFW} shows a density profile
 of our simulated typical rich cluster, CLc,  
and the NFW profile with $c = 4.4$ 
which fits well the numerical result.
Density profiles of dark halos obtained by us agree well with 
the NFW profile in the range 
from the gravitational softening length to $r_{200}$.

\subsection{The differential energy distribution}

We introduce the differential energy distribution, 
$d M/ d\varepsilon$ which gives
the mass of dark matter in the dark halo with binding energy 
between $\varepsilon$ and $\varepsilon + d \varepsilon$,
where $\varepsilon$ is the specific binding energy, 
\begin{eqnarray}
	\varepsilon = \Psi(r) - \frac{1}{2} v^2, 
\end{eqnarray}
and a relative potential, $\Psi \equiv - \Phi + \Phi_0 $.
$\Phi$ is gravitational potential and we choose $\Phi_0$ to
be such that a distribution function, $f$, is $f > 0$ for $\varepsilon > 0$ 
and $f = 0$ for $\varepsilon \le 0$.
In our analysis, $\varepsilon$ is normalized by $GM_{200}/r_{200}$.
And we also introduce the fractional mass distribution as 
the differential energy distribution divided by 
the total mass of the dark halo,
$N(\varepsilon) = d M / d \varepsilon / M$.

\begin{figure}
\plotone{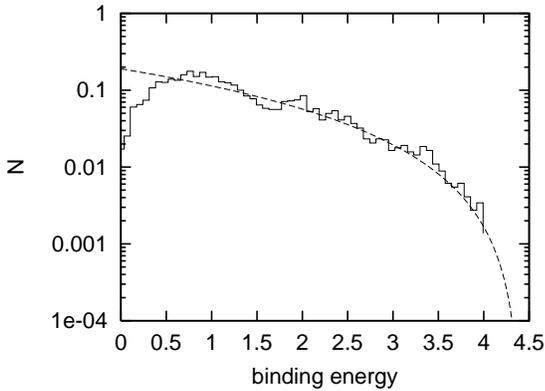}
\caption{Fractional mass distribution of our simulated rich cluster, CLc.
Solid line is our numerical result and dashed line is given by equation (\ref{escort}) with $q = 0.667$ and $\varepsilon_0 = 1.47$.}
\label{fractional_mass_distribution1}
\end{figure}

\placefigure{fractional_mass_distribution1}

Figure \ref{fractional_mass_distribution1} shows the fractional mass distribution of CLc.
In figure \ref{fractional_mass_distribution1}, 
we also show $N(\varepsilon)$ given by
\begin{eqnarray}
	N(\varepsilon) = N_0 \left [ 1 - (1-q) \left( \frac{\varepsilon }
		{\varepsilon_0} \right) \right]^{q/(1-q)},
\label{escort}
\end{eqnarray}
with $q = 0.667$ and $\varepsilon_0 = 1.47$.
Figure \ref{fractional_mass_distribution1} shows that 
equation (\ref{escort}) agrees well with our numerical results in the range of 
$0.5 < \varepsilon < 4$.
There is cut off near $\varepsilon \simeq 4$. 
We find the fractional distribution $N(\varepsilon)$ can be fitted by 
following formula,
for $q \simeq 0.6-0.7$ and 
$\varepsilon_0 \simeq 1.2-1.5 GM_{200}/r_{200}$ for 
rich clusters in our numerical results.

\placetable{table2}

In table 2 we give $N_0$, $q$, 
and $\varepsilon_0$ of our numerical results for rich clusters.

We have shown from our $N$- body simulation that the fractional mass distribution  is also well fitted by the equation (\ref{escort}).
However, near $\varepsilon = 4$ in figure \ref{fractional_mass_distribution1} there are small number of $N$-body 
particles.
We should confirm consistency between 
$N(\varepsilon)$ given by equation (\ref{escort}) and the NFW profile.
We give a fractional mass distribution from 
the NFW profile as follows for the comparison.

We assume that phase-space distribution function $f({\bf x}, {\bf v})$  depends $\varepsilon$. 
At a radius $r$, velocity of a dark matter particle of 
the binding energy, $\varepsilon$, is $v = \sqrt{2(\Psi - \varepsilon)}$.
The density profile may be given as follows (Binney and Tremaine 1987)
\begin{eqnarray}
	\rho (r) = 4 \pi \int_{\Psi(r_g)}^{\Psi(r)} f(\varepsilon) 
		[2 (\Psi - \varepsilon ) ]^{1/2} d \varepsilon, \label{density}
\end{eqnarray}
where $r_g$ is the edge of the dark halo.
From this equation, we may give $f(\varepsilon)$ as
\begin{eqnarray}
	f(\varepsilon) = \frac{1}{\sqrt{8} \pi^2} \frac{d}{d \varepsilon}
	\int_{\varepsilon_{min}}^{\varepsilon} 
	\frac{d \rho/ d \Psi}{[\varepsilon - \Psi ]^{1/2}} d \Psi, \label{fe}
\end{eqnarray}
where $\varepsilon_{min} = \Psi(r_g)$.

Equation (\ref{density}) gives mass $M$ as
\begin{eqnarray}
	M(r) = &&16 \pi^2 \int_0^{r} r^{2} d r \nonumber \\ 
		&& \times \int_0^{\Psi(r)} f(\varepsilon) 
		[2 (\Psi - \varepsilon ) ]^{1/2} d \varepsilon.
\label{mr}
\end{eqnarray}
From equation (\ref{mr}), the differential energy distribution is
\begin{eqnarray}
	\frac{dM(\varepsilon)}{ d \varepsilon} 
	= f(\varepsilon) g(\varepsilon), \label{dmde}
\end{eqnarray}
where
\begin{eqnarray}
	g(\varepsilon) = 16 \pi^2 \int_0^{r_m(\varepsilon)} 
	[2 (\Psi - \varepsilon) ]^{1/2} r^2 dr, \label{ge}
\end{eqnarray}
and $r_m(\varepsilon)$ is maximum radius 
that can reached by a particle of the binding energy $\varepsilon$.

If we assume the density profile is the NFW profile, we get $d M/ d \varepsilon $ from the equations (\ref{fe}), (\ref{dmde}), 
and (\ref{ge}) for the NFW profile. 

\begin{figure}
\plotone{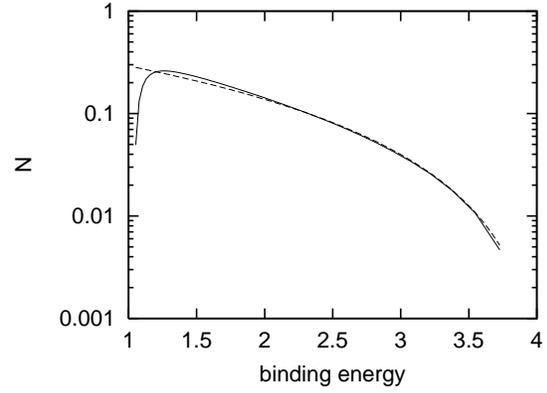}
\caption{Fractional mass distribution of dark halo.
Solid line is caliculated by equations (\ref{fe}), (\ref{dmde}), 
and (\ref{ge}) and dashed line is given by equation (\ref{escort}) 
with $q = 0.66$ and $\varepsilon_0 = 1.4$.}
\label{fractional_mass_distribution2}
\end{figure}

\placefigure{fractional_mass_distribution2}
 
Figure \ref{fractional_mass_distribution2} shows
the fractional mass distribution of NFW obtained in this way and 
$N(\varepsilon)$ given by equation (\ref{escort}).
We show that $N(\varepsilon)$ given by equation (\ref{escort}) is 
consistent with the NFW profile.
We note that the cut off at high binding energy seen in 
figure \ref{fractional_mass_distribution1} is not artifact
due to limitation of our numerical resolutions.

\begin{figure}
\plotone{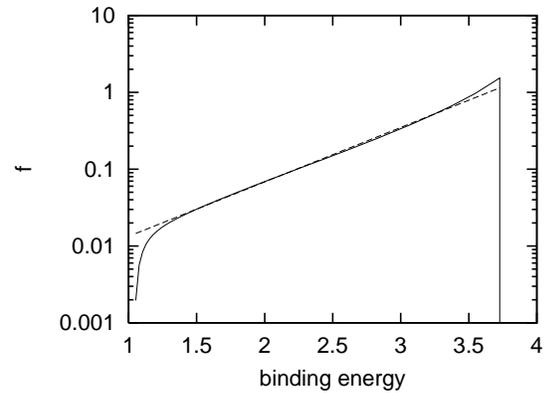}
\caption{$f(\varepsilon)$ of the NFW profile.
Solid line is $f(\varepsilon)$ and dashed line is a fitted exponential 
function.
} 
\label{fe-e}
\end{figure}

\placefigure{fe-e}
 
Figure \ref{fe-e} shows the distribution function of the NFW profile 
obtained by the above method.
Since lower binding energy particles evaporate from 
the dark halo, their fraction become small.
This form of function is similar to the distribution function of King model,
$f_K(\varepsilon) \propto (e^{\varepsilon/\sigma^2} -1)$ (Binney 1982).
On the other hand, there is a peak and sharp cut off at the high binding energy. 
This part corresponds to the central cusp.
Although the distribution function of NFW profile have cut off 
at the high energy, the distribution function of King model 
does not have such cut off.
This is an important difference between them.

\section{The fractional mass distribution, density profile, 
and the distribution function}

We find that the NFW profile satisfies the fractional mass distribution 
given by equation (\ref{escort}) with $q \simeq 0.6-0.7$.
We study how the density profile changes 
when we change the parameter $q$ and $\varepsilon_0$ 
in equation (\ref{escort}).
In this study, we use an iteration method 
as shown in the next subsections.

\subsection{The iteration method}
Binney (1982, and see also Binney and Tremaine 1987) studied 
the phase space structure of galaxies of
which surface brightness is the de Vaucouleurs' $r^{1/4}$ law. 
We apply his method to our study of 
the phase space structure of dark halo with the NFW profile. 
We obtain the density profile and the phase space distribution 
which are consistent with equation (\ref{escort}), 
using the iteration method.
The procedure is as follows.

We assume the dark halo is spherically symmetric. 
From equation (\ref{escort}),
we assume $dM(\varepsilon)/d\varepsilon$ 
given by equation,
\begin{eqnarray}
	\frac{dM (\varepsilon)}{d \varepsilon} = M_0 \left[ 
		1 - (1 - q) \left(\frac{\varepsilon}{\varepsilon_0}\right) 
	\right]^{q/(1-q)},
	\label{escort_m}
\end{eqnarray}
where $M_0 = 1.0$.
We use a trial function of a density profile, 
to calculate a gravitational potential 
in $g_0(\varepsilon)$,
\begin{eqnarray}
	\rho_0(x) = \frac{\rho_c}
	{\left(1 +\left(\frac{x}{x_c}\right)^2 \right)^{3/2}},
\end{eqnarray}
where $x_c = r_c/r_{200} = 0.1$.
This trial function is similar to the $\beta$-model of 
X-ray surface brightness of clusters of galaxies.
We obtain relative potential from this trial density function.
The relative potential is written as
\begin{eqnarray}
	\Psi(x) &= &4 \pi G \left[ 
	\frac{1}{x} \int_0^x \rho_0(x^{\prime} ) x^{\prime^2} d x^{\prime}
	\right.  \nonumber \\
	&&\left. + \int_x^{x_{o}} \rho_0(x^{\prime}) x^{\prime} dx^{\prime}
	\right] + \Phi_{x_{o}}, \label{Psi}
\end{eqnarray}
where $x = r/r_{200}$, and $x_o = 10$ is assumed.
$\Phi_{x_{o}}$ is chosen to
be $f > 0$ for $\varepsilon > 0$ 
and $f = 0$ for $\varepsilon \le 0$.
We obtain $g_0(\varepsilon)$ by equation (\ref{ge}) for $\rho_0$.
From equation (\ref{dmde}),
\begin{eqnarray}
	f_0(\varepsilon) = \frac{dM(\varepsilon)/d \varepsilon}{g_0(\varepsilon)}.
\end{eqnarray}

Next, we calculate new density as,
\begin{eqnarray}
	&&\rho_{i+1}(x) = (1 - \alpha) \rho_i(x)  \nonumber \\
		&&+ \alpha \times 4 \pi \int_{\Psi_i > \varepsilon} 
		f_i(\varepsilon) \sqrt{2(\Psi_i - \varepsilon)} d \varepsilon,
\label{rho_i+1}
\end{eqnarray}
where $i$ is an iterative index and 
$\alpha$ is an arbitrary constant of 
convergence. 
Here we assume $\alpha = 0.5$.
Next we again follow the step from equation (\ref{Psi}) to equation 
(\ref{rho_i+1}) by using $\rho_1$ instead of $\rho_0$ and 
we obtain $\rho_2$.
We continue until $\rho_i$ becomes converge.
In this way, we obtain $\rho$ and $f(\varepsilon)$ which are consistent with equation (\ref{escort_m}).

\subsection{The density profile and the distribution function}

Using the Binney's iteration method, 
we reconstruct mass density profile 
from equation (\ref{escort_m}).

\begin{figure}
\plotone{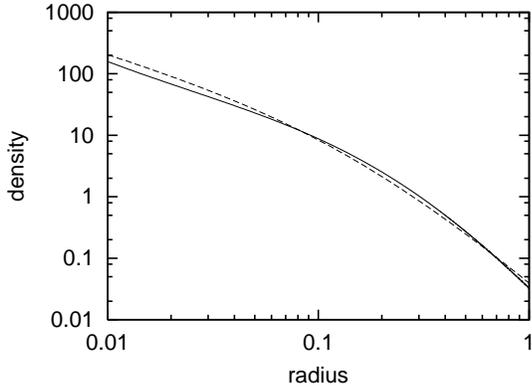}

\caption{Density profile constructed by the iteration method,  assuming the equation (\ref{escort_m}). 
Solid line is our result for $q = 0.67$ and $\varepsilon_0 = 1.4$ 
and dashed line is the NFW profile of $c = 6.59$.
Radial distance is normalized by $r_{200}$.}
\label{constructed_density1}
\end{figure}

\placefigure{constructed_density1}

We show our result for $q=0.67$ and $\varepsilon = 1.4$ in figure 
\ref{constructed_density1}.
A solid line is given by the iteration method and 
a dashed line is the NFW profile.
We confirm that $dM(\varepsilon)/d \varepsilon$ characterizes 
well the NFW profile. 

\begin{figure}
\plotone{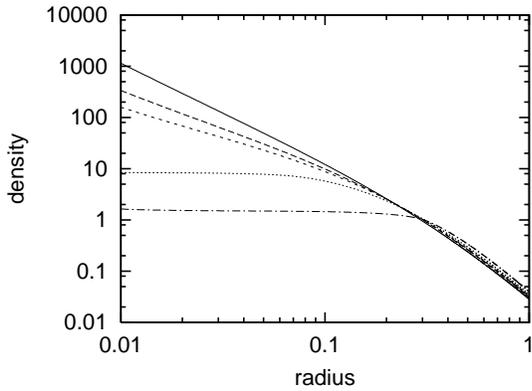}
\caption{
Density profiles for various $q$.
Dot-dashed, dotted, short-dashed, dashed, and solid lines are for 
$q = 0.25, 0.5, 0.67, 0.75, $ and $1.0$, respectively.
Density profile with smaller $q$ has a flat core.
On the other hand, one approaches $\rho \propto r^{-2}$
in $r < 0.1$ for $q \longrightarrow 1$.}
\label{q_density_relation}
\end{figure}

\placefigure{q_density_relation}

In figure \ref{q_density_relation}, 
we show density profiles for different $q$ but $\varepsilon = 1.4$.
Smaller $q$ (e.g. $q = 0.5$) results in shallower core 
in the inner region.
On the other hand, larger $q$ ($q > 0.67 $) 
makes a cusp steeper than the NFW profile, density profile
approaches $\rho \propto r^{-2}$ 
in the inner part, for $q \longrightarrow 1$.

\begin{figure}
\plotone{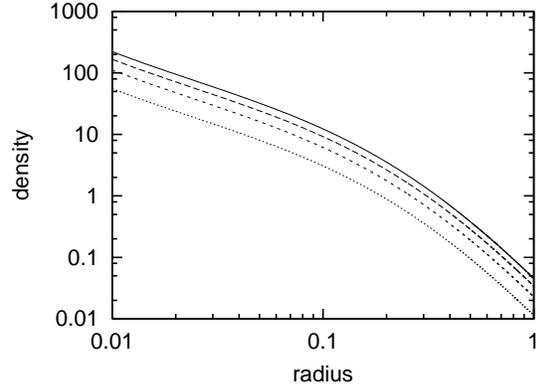}
\caption{
Density profiles for various $\varepsilon$.
Dotted, short-dashed, dashed, and solid lines are for 
$\varepsilon_0 = 0.5, 1.0, 1.5, $ and $2.0$, respectively.
Density profiles are self-similar. 
However, their normalizations are different.}
\label{e_density_relation}
\end{figure}

\placefigure{e_density_relation}

For various values of $\varepsilon_0$, 
the density profiles are similar to the NFW for $q = 0.6$ 
as shown in figure \ref{e_density_relation}.
Absolute value of the density depends on $\varepsilon_0$.
Therefore, the slope of the cusp depends on only $q$, not $\varepsilon_0$.

\begin{figure}
\plotone{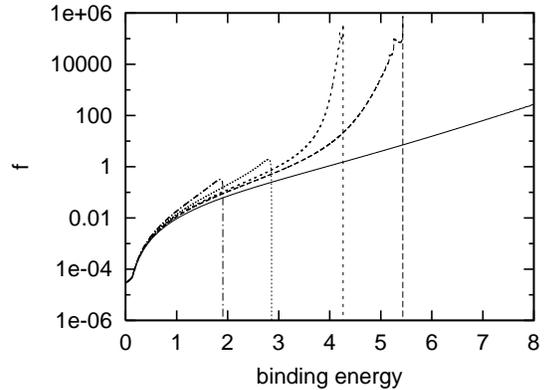}
\caption{We show $f(\varepsilon)$ for various $q$ ; 
Dot-dashed, dotted, short-dashed, dashed, and solid lines are for 
$q = 0.25, 0.5, 0.67, 0.75, $ and $1.0$, respectively.}
\label{q_dependence_f}
\end{figure}

\placefigure{q_dependence_f}

Figure \ref{q_dependence_f} shows 
the distribution function, $f$, obtained by the iteration method for 
various values of $q$ .
These curves show the same dependence on $\varepsilon$ in $0 < \varepsilon <1$.
Peak values of $f$ are different each other.
We also show the Boltzmannian distribution for comparison 
in figure \ref{q_dependence_f}.
For large $q$, peak value of $f(\varepsilon)$ and 
the maximum binding energy of the distribution become large.
We have shown that the density profile with large $q$ have the steep cusp.
Therefore, sharp peak of $f(\varepsilon)$ corresponds to the steep cusp.

\begin{figure}
\plotone{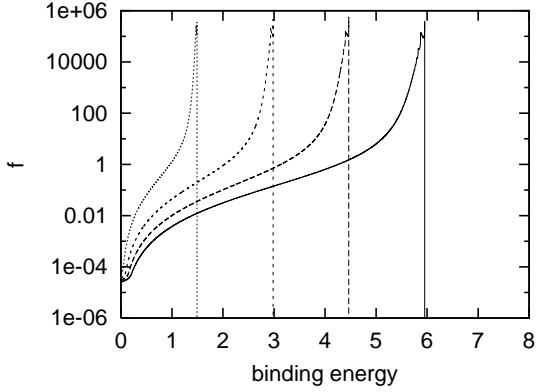}
\caption{We show $f(\varepsilon)$ for various $\varepsilon_0$;
Dotted, short-dashed, dashed, and solid lines are for 
$\varepsilon_0 = 0.5, 1.0, 1.5$, and $2.0$, respectively.}
\label{e_dependence_f}
\end{figure}

\placefigure{e_dependence_f}

Figure \ref{e_dependence_f} shows $f$ obtained by the iteration method
for same $q$ but various $\varepsilon_0$.
These curves do not show the same dependence on binding energy.
Height of peaks of these curves is constant.

\section{Summary and Discussion}
We analyze the universal density profile of dark halo proposed by NFW and its
differential energy distribution.
Our main results are summarized as follows.

\begin{enumerate}
\item
We study
the fractional mass function $N(\varepsilon)$ for dark halo obtained 
by our numerical simulation and find its analytical formula 
which is the equation (\ref{escort}).
\item We show that the NFW profile is 
given by the equation (\ref{escort_m}).
\item We show that the slope of the cusp in the density profile changes with a value of the parameter $q$ in the analytical formula.
\end{enumerate}

We can regard that $N(\varepsilon)$ shows the statistical property of the NFW profile.
If the NFW profile is universal, $q = 0.6 - 0.7$ in equation (\ref{escort}).
Different $q$ makes slope of a cusp different.
Since $q$ plays an important role, we should make clear 
what physical process determines $q$. 
Recent high resolution numerical simulation 
(Okamoto \& Habe 1999, 2000) shows the steeper cusp,
$\rho \propto r^{-1.5}$, than the the NFW profile.
This profile corresponds to $q = 0.75 - 0.8$.
Isothermal profile, $\rho \propto r^{-2}$, corresponds to $q = 1$.

We study $f(\varepsilon)$ for the NFW profile. 
The formula of this is not isothermal one nor 
the King formula $f_K \propto e^{\varepsilon/\sigma^2} -1$.
$f(\varepsilon)$ for the NFW profile have the energy cut off 
at the high end of $\varepsilon$.
We should study the reason why $f(\varepsilon)$ has such a form.
Lynden-Bell (1967) studied distribution function $f(\varepsilon)$ of a 
virialized system.
Maximizing the Boltzmann entropy of the system, 
resulting distribution is isothermal profile, $ \rho \propto r^{-2}$.
In this case the system has infinite extend and infinite mass.
This is not realistic for astronomical objects.
Cosmological simulations have shown that 
galaxies and clusters of galaxies formed in these simulations have 
more rapid radial decline than isothermal in the outer part.

We note that the form of the distribution function 
relates with the cusp profile.
For $q < 1$, there are the maximum binding energy 
$\varepsilon_{{\rm max}}$ and the sharp peak in $f(\varepsilon)$ 
at $\varepsilon_{{\rm max}}$.
Since there is $\varepsilon_{{\rm max}}$, 
phase space volume occupied by dark matter is limited 
than the isothermal profile.
This may be the reason of 
the cusp profile different from $\rho \propto r^{-2}$.
For $q < 1$, the distribution function $f(\varepsilon)$ changes with 
$q$ values as shown in figure 8.

We note that the form of equations 
(\ref{escort}) and (\ref{escort_m}) are 
similar to the Tsallis' escort distribution,
\begin{eqnarray}
	\frac{P(E,T^{\prime})}{P(0,T^{\prime})} = 
		\left[ 1 - ( 1- q) \frac{E}{T^{\prime}} \right]^{q/(1-q)},
\end{eqnarray}
where $T^{\prime}$ is temperature parameter 
and $q$ is entropic index (Tsallis, Mendes, \& Plastino, 1998).
Tsallis' non-extensive generalized statistics (Tsallis 1988) is 
paid attention in the area of statistics of a multi-fractal system. 
In non-extensive system 
(long-range microscopy memory, long range forces, 
fractral space time) the following generalized entropy has been proposed :
\begin{eqnarray}
	S_q = k \frac{1 - \sum_i p_i^q}{q-1} \ \ \ 
	\left( \sum_i = p_i = 1; \ \ \ q \in \Re \right),
\end{eqnarray}
where $k$ is a positive constant.
Optimization of $S_q$ yields, for the canonical ensemble,
\begin{eqnarray}
	&&p_i = Z_q^{-1} [1 - (1-q) \varepsilon_i/T^{\prime} ]^{1/(1-q)}, \\
	&&Z_q \equiv \sum_i [1 - (1-q) \varepsilon_i/T^{\prime} ]^{1/(1-q)}
\end{eqnarray}
and, when $q \longrightarrow 1$, 
the Boltzmann-Gibbs result is recovered.
In this statistics an expected value of any physical variable is given by 
the Tsallis' escort distribution :
\begin{eqnarray}
	\left< A \right>_q = \frac{ \sum_i p_i^q A_i}{\sum_j p_j^q},
\end{eqnarray}
where $\{A_i\}$ are the eigenvalues of an arbitrary observable $A$.
The escort distribution, $P_i = p_i^q/\sum_j p_j^q$, is 
similar to the form of equation (\ref{escort}) and (\ref{escort_m}).

It is expected that long range interaction makes 
the system non-extensive.
Our case may be the one of this cases.
Our results suggests that differential energy distribution of 
collisionless particles is described by the escort distribution 
of the Tsallis statistics.
This may indicate that energetic process of collisionless particles must be stochastic process.

In the Tsallis' escort distribution function, 
there is the maximum value of $\varepsilon$ 
for $p_i > 0$ for $0 < q < 1$.
This case is called superextensive.
The shallower cusp profile in the the NFW profile shows 
the superextensive property of the dark matter distribution.

We show that the NFW profile corresponds to $q = 0.6-0.7$.
Lavagno {\it et al.} (1998) have recently shown 
that fraction of peculiar velocity of
 cluster of galaxies (Bahcall \& Oh 1996) is well explained by 
the Tsallis escort integral,
\begin{eqnarray}
	P(> v) = \frac{\int_v^{v_{max}} (1 - (1-q)(v/v_0)^2)^{q/(1-q)}dv}
	{\int_0^{v_{max}} (1 - (1-q)(v/v_0)^2)^{q/(1-q)}dv}.
\end{eqnarray}
They obtained $q=0.23$ to fit the fraction of peculiar velocity of cluster of galaxies and is smaller than in our case.
There is a conjecture that $q$ of system approaches unity 
when the system proceeds relaxation (Tsallis 1999).
Our results in which $q = 0.6-0.7$ 
are consistent with this conjecture,
since dark matter distribution in a cluster of galaxies is 
more relaxed system than 
the large scale motion of clusters of galaxies. 
The value of $q$ must be related with 
the degree of relaxation of collisionless particles.
$q = 1$ corresponds to isothermal distribution in the Tsallis statistics.
The gravitational system like clusters of galaxies 
with the universal profile may be non-extensive 
because they formed recently.

We should study the reason why dark halo has the value of $q = 0.6-0.7$ in the hierarchical clustering scenarios.
It is interesting to study the differential energy distribution of 
self-gravitational system formed in a circumstance without 
hierarchical clustering to make clear mechanism what determines $q$ of 
the gravitational system.

\acknowledgments
We would like to thank Sumiyoshi Abe, Masayuki Fujimoto, and Seiichi Yachi for helpful comments and discussion.

\begin{table}
\caption{The physical values of our simulated clusters}
\begin{tabular}{lccc}
\hline \hline
& Mass & $T$ & $c$  \\
&($10^{15}M_{\odot}$) & ($10^8$ K) \\ \hline
CLa & 3.05 & 0.953 & 2.50 \\
CLb & 2.80 & 0.890 & 2.72 \\
CLc ($z = 0$) & 3.36 & 1.29 & 4.41 \\
CLc ($z = 0.25$) & 1.60 & 0.796 & 2.31  \\
CLc ($z = 0.5$) & 0.831 & 0.540 & 2.67  \\ 
CLd & 3.71& 1.57 & 6.34  \\
CLe & 2.82& 1.19 & 4.65  \\
CLf & 2.31& 0.654 & 2.75  \\ \hline
\end{tabular}
\tablecomments{
Each value is calculated at redshift $z = 0$, except for CLc.
}
\label{table1}
\end{table}

\begin{table}
\caption{$N_0$, $q$, and $\varepsilon_0$ of rich clusters in 
our numerical results}
\begin{tabular}{lccc}
\hline \hline
&  $N_0$ & $q$ & $\varepsilon_0$ \\ \hline
CLa &  0.309 & 0.612 & 1.27\\
CLb &  0.253 & 0.680 & 1.21\\
CLc ($z = 0$) &  0.190 & 0.667 & 1.47 \\
CLc ($z = 0.25$) &  0.230 & 0.606 & 1.05 \\
CLc ($z = 0.5$) &  0.305 & 0.600 & 0.672 \\ 
CLd & 0.276 & 0.699 & 1.26 \\
CLe & 0.117 & 0.503 & 2.03 \\
CLf &  0.258 & 0.580 & 1.41 \\ \hline
\end{tabular}

\tablecomments{
Each value is calculated at redshift $z = 0$, except for CLc.
}
\label{table2}
\end{table}

\end{document}